\documentclass[fleqn,12pt]{wlscirep}

\usepackage[utf8]{inputenc}
\usepackage[T1]{fontenc}
\usepackage[toc,page]{appendix}
\usepackage{amsmath}

\usepackage{authblk} 
\usepackage[numbers,sort&compress]{natbib}

\usepackage[parfill]{parskip}
\usepackage{color}
\definecolor{mygreen}{rgb}{0.1,0.6,0.2}

\title{Mathematical model bridges disparate timescales of lifelong learning}

\author[1,*]{Mingzhen Lu}
\author[2, *]{Tyler Marghetis}
\author[1, *]{Vicky Chuqiao Yang}

\affil[1]{Santa Fe institute, 1399 Hyde Park Road, Santa Fe NM 87501, USA. }
\affil[2]{University of California, Merced, 5200 Lake Rd, Merced, CA 95343, USA. }
\affil[*]{All authors contributed equally and are co-corresponding}

\usepackage{natbib}
\usepackage{graphicx}

\begin{abstract}

Lifelong learning occurs on timescales ranging from minutes to decades. People can lose themselves in a new skill, practicing for hours until exhausted. And they can pursue mastery over days or decades, perhaps abandoning old skills entirely to seek out new challenges. A full understanding of learning requires an account that integrates these timescales. Here, we present a minimal quantitative model that unifies the nested timescales of learning. Our dynamical model recovers classic accounts of skill acquisition, and describes how learning emerges from moment-to-moment dynamics of motivation, fatigue, and work, while also situated within longer-term dynamics of skill selection, mastery, and abandonment. We apply this model to explore the benefits and pitfalls of a variety of training regimes and to characterize individual differences in motivation and skill development. Our model connects previously disparate timescales---and the subdisciplines that typically study each timescale in isolation---to offer a unified account of the timecourse of skill acquisition.

\end{abstract}

\begin{document}

\maketitle
\thispagestyle{empty}





\section*{Significance}

Learning is a complex process that spans timescales from moments to decades. Previous work has identified regularities on long timescales (e.g., diminishing returns on effort) and short timescales (e.g., flow states). Here we develop a first-principles model that integrates these timescales. We show how long-timescale learning can emerge from short-timescale oscillations in motivation, fatigue, and engagement; and conversely how these short-timescale dynamics are shaped by long-timescale changes in skill and task difficulty. We also demonstrate how individual differences in sources of motivation can lead to different rates of skill acquisition. This minimal model of sustained learning suggests ways to help people persevere through challenges, cultivate personal excellence, and optimize life-long development.

\textbf{Keywords}
learning, timescales, multiscale model, dynamical system, skill development

\section{Introduction}


\subsection{Lifelong Learning}
Humans are perpetual learners. Throughout their lives, most individuals will acquire one or more languages, learn motor skills such as riding a bicycle or driving a car, and master complex abilities for work or pleasure \cite{faure1972learning}. Ongoing learning is a through-line of human life, no matter the skill level, from children at play to Picasso at his peak. 

This ongoing process of learning occurs on multiple timescales \cite{newell2001time, huys2004multiple, smith2003development}. Over the long timecourse of mastering a skill, learning may appear continuous and gradual as one progresses from novice to mastery. But this slow process consists of countless shorter, discrete periods of engagement, lasting from mere moments to entire days. An individual may start working but get bored. They may rest. They may eventually master the skill, or give up, or switch focus to another skill entirely. In short, learning occurs on multiple nested timescales, from the moment-to-moment dynamics of motivation and fatigue, to briefer episodes of sustained effort, to the slow crawl toward mastery. 

A full understanding of learning requires an account that integrates these timescales \cite{smith2003development}. Existing accounts, however, have typically focused on single timescales. Here, we present a minimal model that unifies the nested timescales of lifelong learning. The model captures the relations between the dynamics of motivation, fatigue, and sustained engagement on the shorter timescale of individual periods of work (e.g., minutes or hours), as well as the dynamics of learning and performance on the longer timescale of mastery (e.g., days or years).

Below, we briefly review the longer and shorter timescales of learning. We then introduce our formal model, which unifies these timescales. 

\subsection{Long timescale}

An often-repeated clich\'{e} holds that mastery requires ``10,000 hours''---that is, two years of constant, daily effort, breaking only to sleep at night. While overly simplifying, this saying captures the fact that real mastery requires sustained engagement over long periods \cite{ericsson1996expert}.

The dynamics of learning on this timescale are typically described using so-called ``learning curves'' or ``progress curves'' \cite{estes1956problem, newell1981mechanisms,anderson1994acquisition, ericsson1996expert}. These represent the gradual increase in skill over time, where skill is typically measured at the level of an entire session of work or practice: the final score in a game, the best performance in some athletic skill within a training session, mean typing speed within a testing period, etc. Progress curves typically show diminishing returns, with the greatest rate of learning when skill is far from mastery, and a gradually diminishing rate of learning as mastery is approached (Fig.~\ref{fig:conceptual}-A, inset) \cite{newell1981mechanisms, ericsson1996expert, anderson1994acquisition, estes1956problem, Stafford2014}. Empirical progress curves can be described using power or exponential functions, although there is debate about the best-fitting functional form \cite{gallistel2004learning, haider2002aggregated, heathcote2000power, myung2000toward, donner2015piecewise}.

Indeed, more recent work has found that, at the individual level, learning often improves in distinct stages (Fig.~\ref{fig:conceptual}-A), reflected in plateaus in the learning curve \cite{gallistel2004learning, gray2017plateaus, van1992stagewise}. These plateaus are often periods where the learner has mastered a particular strategy that is successful but sub-optimal. Having mastered this strategy, the learner may then pursue a better one, which might produce worse performance at first but ultimately better performance as the new strategy is mastered (Fig.~\ref{fig:conceptual}-B) \cite{gray2017plateaus}. Thus, on the slow timescale of skill mastery, individuals typically exhibit learning with diminishing returns, and may exhibit repeated plateaus, followed by brief dips in performance and increased learning. 

\subsection{Short timescale}
The slow process of skill mastery, however, is built out of many brief sessions of practice and learning. People choose to start working, and then persist in that effort until they give up or require rest---and then, having rested, they may choose to start once more. When people start and stop is often a function of their motivation \cite{katzell1990work} and fatigue \cite{konz1998work}.\footnote{Exceptions include cases where people are forced to work by some external entity, such as a teacher, coach, or employer.} While the development of skill and performance are typically measured at the level of entire work-sessions, people are nevertheless learning within each session. On this fast timescale of moment-to-moment engagement, motivation and fatigue may fluctuate, skills may creep ever upward, and accomplishments may accumulate slowly. Long-term mastery, therefore, emerges from the the dynamics of short-timescale engagement and learning.

\begin{figure}
    \centering
    \includegraphics[width = 0.65\textwidth]{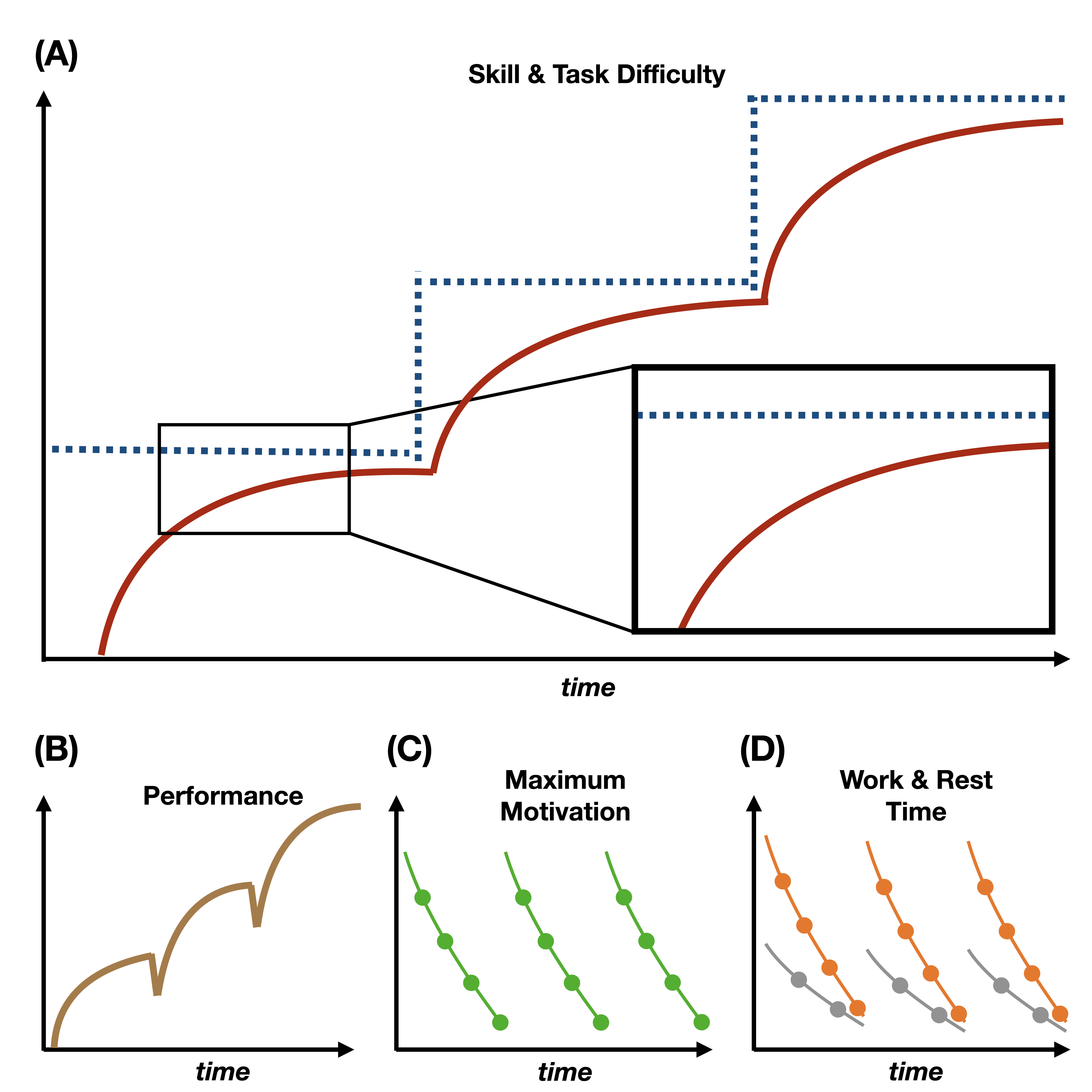}
    \caption{\textbf{Dynamics of learning, performance, and sustained engagement.} (A) Skill development (solid red) typically exhibits diminishing returns on practice as one approaches mastery of a task. Over time, one can switch to more challenging tasks (dashed blue line), thus creating opportunities for continued skill development. Zoomed inset shows a typical learning curve with diminishing returns over time. (B) Performance increases with practice but can reach plateaus. When one switches to a more challenging task, there can be a momentary dip in performance before performance outstrips the previous plateau. (C) For people who are motivated by skill development, motivation reaches a peak when the task is sufficiently difficult to maximize learning. As the gap shrinks between skill and task difficulty, motivation decreases. Switching to a more challenging task can increase motivation. (D) Individuals motivated by skill development work for long periods, and require relatively little rest, when a task is sufficiently challenging. As the gap shrinks between skill and task difficulty, they work for shorter and shorter periods.
}
    \label{fig:conceptual}
\end{figure}

One influential phenomenon at this short timescale is the experience of flow. Experiences of flow consist of a cluster of features, including higher motivation and an ability to work for long periods without interruption \cite{csikszentmihalyi1975flowing,csikszentmihalyi1990flow}. People experiencing flow often report losing track of time and working unceasingly without food or rest. Classic research has found that a person is most likely to experience flow when the task is sufficiently challenging, relative to their current skill. When somebody is ``in the zone,'' as flow is sometimes described, they will experience heightened motivation (Fig.~\ref{fig:conceptual}-C) and will persist in practicing a difficult skill despite fatigue (Fig.~\ref{fig:conceptual}-D).
    
\subsection{Unifying timescales}
A full understanding of life-long learning and sustained engagement requires an understanding of both long and short timescales \cite{smith2003development}. At present, however, we lack a unifying framework, one that allows situating the short timescale of single work sessions within the context of long-timescale skill development.

This is a problem because changes on one timescale shape changes on the other. Lifelong learning emerges from the moment-to-moment dynamics of motivation, fatigue, skill development, and performance. The decision to work or rest, meanwhile, is a function of fatigue and motivation, but motivation is shaped by long-timescale changes in skill and in the task being learned. The dynamics of each timescale only make sense in the light of the other.

Encouraged by successful models of nested timescales in other biological and social systems \cite{hastings2010timescales, hastings2016timescales, flack2012multiple, wagner2021modeling}, here we develop a first-principle mathematical model that unifies the nested timescales of learning. By bridging between levels --- the long timescale of mastery, and the short timescale of effort and engagement --- the model connects phenomena that have traditionally been investigated by distinct research traditions. 

In what follows, we introduce the formal model. We then describe the behavior of this model on multiple timescales, in the context of different training regimes (e.g., continuously increasing task difficulty vs. discrete jumps in difficulty), and for individuals who differ in their motivation.

\section{Summary of the Mathematical Model}
We present a first-principle model that integrates the multiple timescales of learning, from moment-to-moment engagement to life-long growth and skill acquisition. We take a dynamical system approach, where our model consists of a system of ordinary differential equations. The dynamical system approach has proven useful in elegantly explaining complex phenomena in human systems such as motor coordination \cite{haken1985theoretical, richardson2007rocking, turvey1990coordination}, cognitive development \cite{smith2003development, van1992stagewise}, left-handedness in a minority of the population \cite{abrams2012model}, and the polarization of political parties \cite{yang2020us, leonard2021nonlinear}. 

\begin{figure}
    \centering
    \includegraphics[width = 1\textwidth]{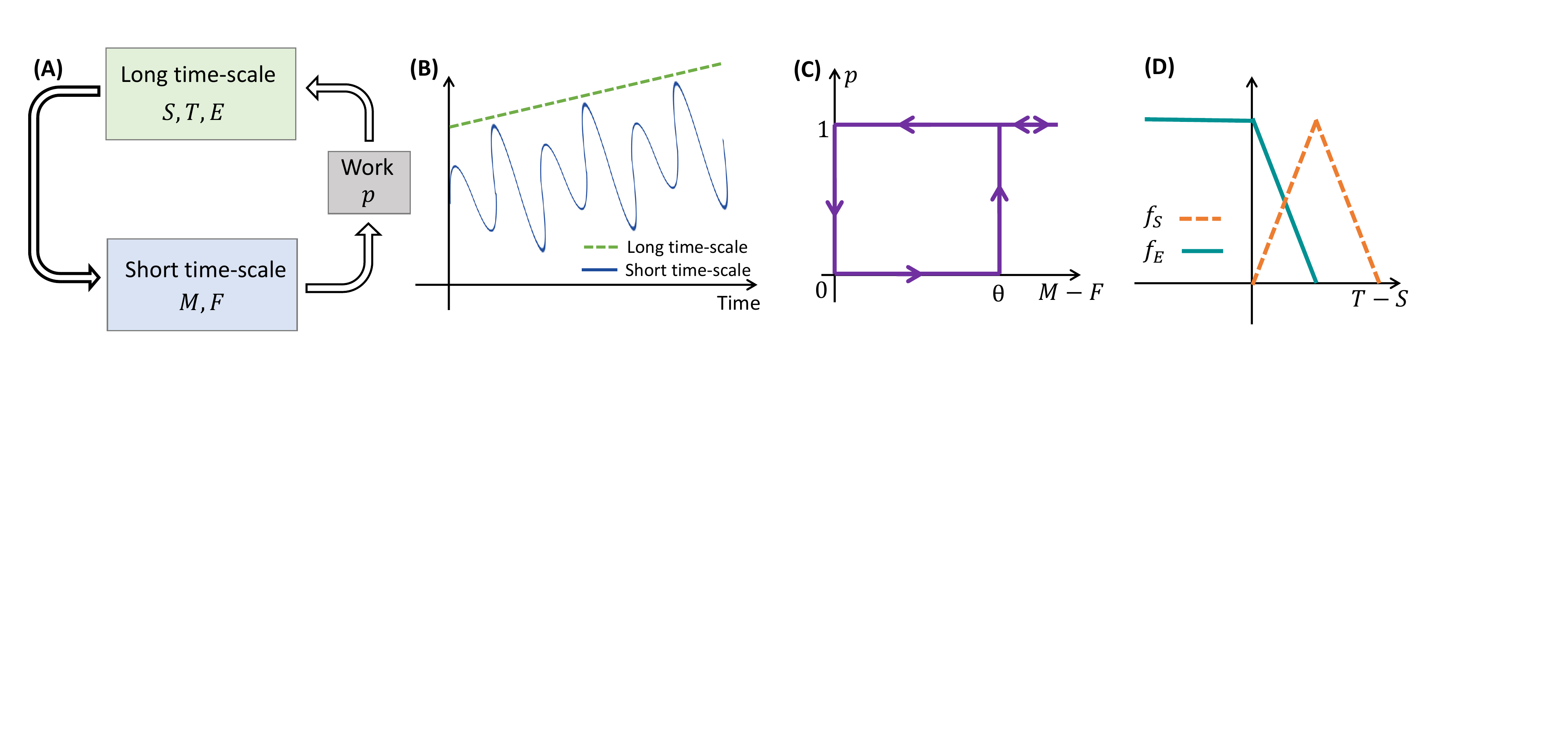}
    \caption{\textbf{Conceptual illustration of the modeling framework and key variables.} (A) Short timescale variables (Motivation, $M$, and Fatigue, $F$) determine whether an individual decides to work, $p$. When an individual works, they can improve their skill $S$, and task execution, $E$, although these change more slowly. Task difficulty, $T$, can change over time or remain constant. These long-timescale variables, in turn, influence the dynamics of short-timescale variables. (B) Long- and short-timescale variables exhibit different dynamics. Long timescale variables (e.g., skill) tend to change slowly . Short timescale variables (e.g., motivation) tend to oscillate. (C) The decision to work ($p=1$) or rest ($p=0$) is a function of the difference between motivation and fatigue ($M-F$). Starting to work requires sufficient motivation relative to fatigue (i.e., $M-F>\theta$). Work stops when fatigue overtakes motivation (i.e., $M-F \leq 0$). The additional effort required to start working, $\theta$, induces hysteresis. (D) Skill $S$ and execution $E$ change more slowly, and their dynamics depend on the relative task difficulty (i.e., $T-S$). The rate of change of skill, $\frac{dS}{dt}$, is described by function $f_S$: learning stops if the task is too easy (i.e,. $T-S$ is negative) or too difficult (i.e., $T-S$ is too large), and reaches a maximum when the task is difficult but not too difficult. One's execution of the task, $E$, is described by function $f_E$, which is at maximum when task is very easy (relative task difficulty is negative or zero), declines linearly as relative task difficulty increases, and stops entirely when relative task difficulty is over a certain threshold. 
    }
    \label{fig:model}
\end{figure}

Here we summarize the model, while the Methods section contains the detailed derivation of the dynamical model. Figure~\ref{fig:model} illustrates the mathematical model's conceptual framework and the relationships between key variables. The mathematical model contains variables that change on both short and long timescales, with the different timescales connected by the individual's decision to work or rest (Fig.~\ref{fig:model}-A). 

Key long-timescale variables are $S(t)$, the skill level of an individual at time $t$; $E(t)$ the rate of execution, or how much work an individual is getting done at time $t$; and $T(t)$, the task difficulty. Individuals increase their skill (i.e., learn) whenever they are working on tasks that are more difficult than their current skill level, but neither too easy nor too difficult (Fig.~\ref{fig:model}-D, dashed line). Execution is maximized when skill is greater than task difficulty, and drops off as the task difficulty becomes increasingly greater than one's skill (Fig.~\ref{fig:model}-D, solid line). Since performance on a task reflects both skill and execution, we calculate performance as the product of $S(t)$ and $E(t)$.

The variables governing moment-to-moment behavior change much more rapidly (Fig.~\ref{fig:model}-B). Key short-timescale variables are the level of motivation, $M$, and fatigue, $F$. The rate of change in motivation is affected by two components. First, motivation increases when an individual is learning or executing a task. Second, in the absence of other influences, motivation will restore to a baseline. The rate of change of fatigue is similarly affected by two components: fatigue increases when one is working, and restores to a baseline when one is not. The individual's decision to work or rest at time $t$ depends on the difference between motivation and fatigue (Fig.~\ref{fig:model}-C). 

We solve for the time trajectories of these variables when task difficulty, $T$, remains constant, and under a variety of regimes of increasing task difficulty (e.g., continuous increase; stepwise increasing in response to improved skill). On the long timescale, we investigate the rate of skill acquisition and execution of tasks. On the short timescale, we investigate the duration of work and rest periods, as well as the dynamics of motivation and fatigue. We also explore differences in skill development between individuals who are motivated by learning and those who are motivated by performance.

\section{Results}
\subsection{Diminishing return in skill and performance under constant task difficulty}
We first analyze a baseline scenario where an individual experiences constant task difficulty over time, illustrated in Fig.~\ref{fig:result_const_T}-A (hereafter \emph{constant scenario}). Real-world examples of this constant scenario are ubiquitous, at least within delimited time periods (e.g., lifting a certain weight, playing a card game against a fixed computer opponent). Note that this scenario is analogous to the regime illustrated in the inset of Fig.~\ref{fig:conceptual}-A.

In this constant scenario, our model predicts that long-timescale skill development reaches a plateau of both skill and performance, as shown in Fig.~\ref{fig:result_const_T}-A and B. This model prediction is consistent with the classic empirical observation of diminishing returns in skill learning when task difficulty is held constant \cite{estes1956problem,  newell1981mechanisms, ericsson1996expert, anderson1994acquisition}. 

These long-timescale patterns are accompanied by changes in short-timescale dynamics. The model predicts that, as skill improves, both peak motivation and peak fatigue within working sessions will decline monotonically (Fig.~\ref{fig:result_const_T}-C, D). Moreover, work sessions last longer --- that is, engagement is more sustained --- when the task difficulty is suitably higher than the skill level (Fig.~\ref{fig:result_const_T}-E). Approaching the plateau, one becomes less and less motivated. Consequently, one tends to work for shorter and shorter periods, suggesting an increasing degree of boredom.

\subsection{Sustained skill learning under stepwise increases in task difficulty}
In real-world settings, individuals tend to engage in increasingly difficult tasks throughout the long-term learning process. To capture this, we evaluated a scenarios where task difficulty increases in discrete jumps whenever an individual's skill catches up with task difficulty, as shown in Fig.~\ref{fig:result_3_panel}-A (hereafter \emph{stepwise scenario}). Real-world situations corresponding to this situation include strength training with incrementally heavier weights, attempting a new level in a video game, or skipping a grade in school. We set task difficulty to increase by a fixed increment of $0.8$ whenever the gap between an individual's skill and the demands of the task (i.e., $T-S$) is smaller than a threshold, $\epsilon$.  

Within each period of constant task difficulty, behavior was similar to that found in the constant scenario, with plateaus in skill and performance as skill approached task difficulty. With repeated, discrete increases in task difficulty, however, both skill and performance surpassed these plateaus (Fig.~\ref{fig:result_3_panel}-A1, A2). 

Our model predicts that when the task difficulty is raised adaptively, one's performance exhibits plateaus, dips, and leaps, as shown in Fig.~\ref{fig:result_3_panel}-B. Whenever task difficulty increases after reaching a plateau, performance first dips, then increases to exceed the previous plateau, in line with empirical studies of performance plateaus \cite{donner2015piecewise}.

Similar to the model's predictions in the constant scenario, each episode of task difficulty level starts with a spike in peak motivation, a spike in peak fatigue, and longer durations of working periods. This suggests intense engagement as one first attempts to master a higher task level. Peak motivation, peak fatigue, and working period duration all decline as skill improves and approaches the target task difficulty, essentially recapitulating what we find in the constant scenario (Fig.~\ref{fig:result_3_panel}-A3, A4, A5).

\subsection{Skill learning under continuous increases in task difficulty}
\begin{figure}
    \centering
    \includegraphics[width = 1\textwidth]{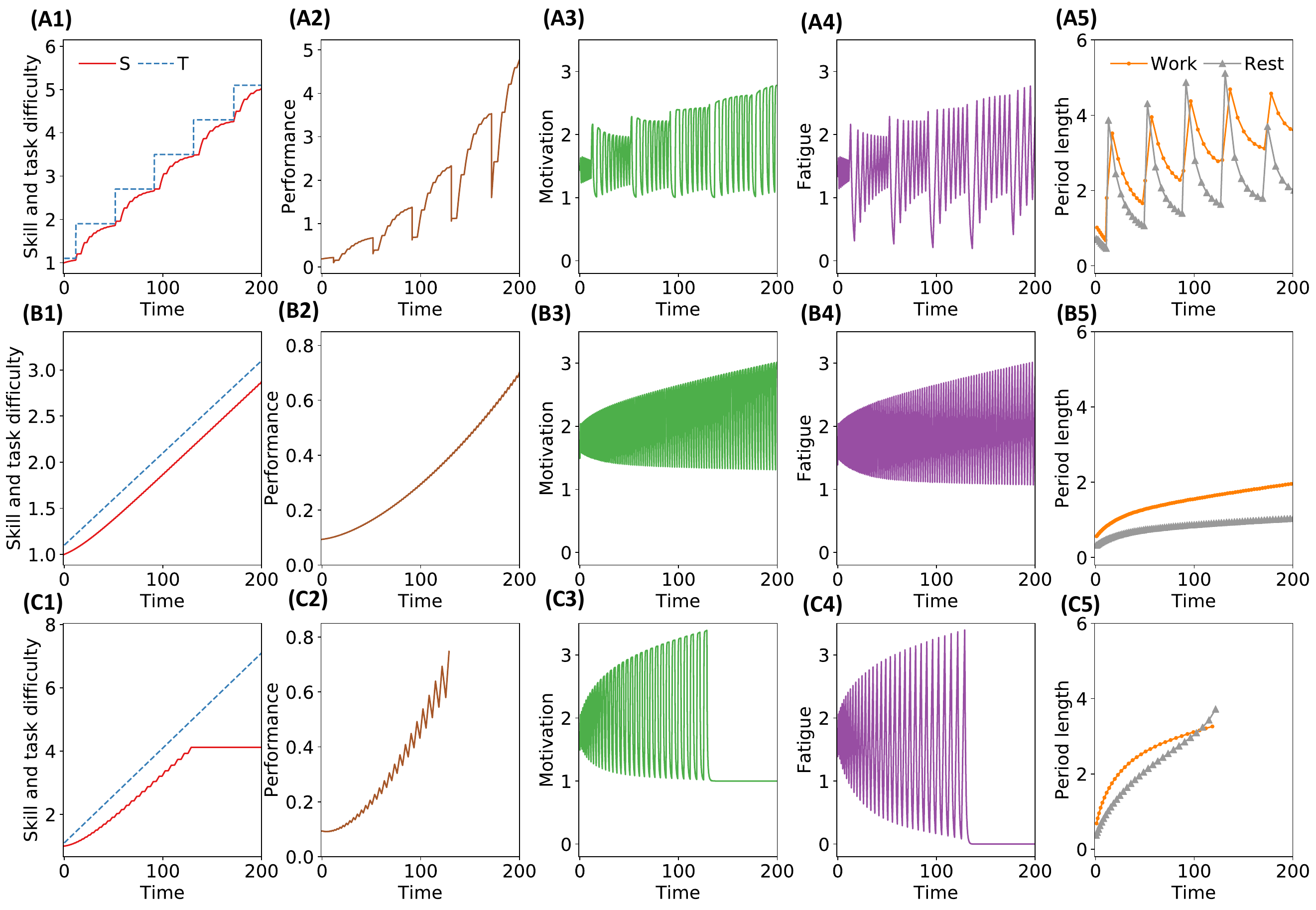}
    \caption{\textbf{Short- and long-timescale dynamics of three distinct learning regimes.} Model predictions for (A) stepwise adaptive task difficulty; (B) linearly increasing task difficulty at a slow pace, and the learner adapt to increasing task difficulty by improving skill; (C) linearly increasing task difficulty at a fast pace, learner cannot keep up with the fast pace and stop working. In all panels, parameter $w = 0.8$.  
    }
    \label{fig:result_3_panel}
\end{figure}

To incorporate activities that are not characterized by stepwise change of difficulty, we also examine a scenario where task becomes increasingly more challenging in a continuous manner (e.g., a gradually increasing speed of running and cycling (Fig.~\ref{fig:result_3_panel}-B and C).

We observe a qualitatively similar pattern in skill learning and performance compared to that in the stepwise scenario---both scenarios induce sustained increases in skill and performance over the long timescale, with the continuous scenario producing smoother changes. 

\begin{figure}
    \centering
    \includegraphics[width = 1\textwidth]{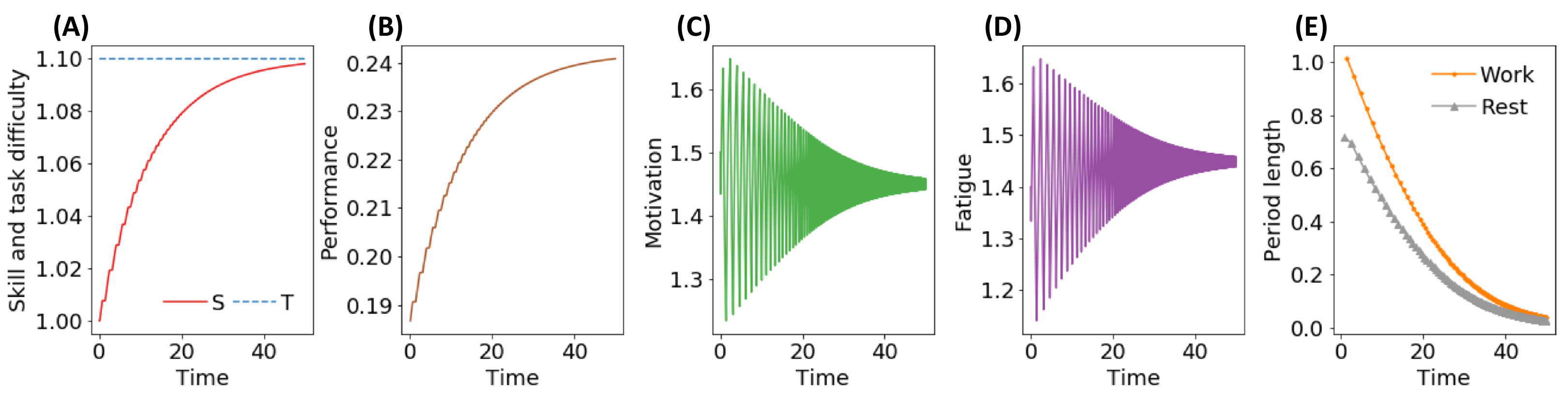}
    \caption{\textbf{Constant task difficulty leads to plateauing skill and performance on the long timescale (A,B), and reduced motivation and working periods on the short timescale (C--E)}. (A) Skill (red solid line) approaches task difficulty (blue dashed line), and eventually plateau. (B) Performance of an individual ($S(t)E(t)$), increases over time but eventually plateau. (C-D) Motivation and fatigue oscillate across recurring working sessions, where individual enter work session when motivation is sufficiently larger than fatigue and exit work session when motivation drops below fatigue (Eq.3). Peak motivation value of each work session diminishes as one's skill and performance plateau.  (E) The durations of individual's working period and rest period both decline over time, consistent with the decreasing motivation. }
    \label{fig:result_const_T}
\end{figure}

When the rate of increase in task difficulty is not overwhelming, there is a gradual increase in peak motivation, peak fatigue, and the duration of working periods (Fig.~\ref{fig:result_3_panel}-B3, B4, B5). This indicates increased levels of engagement. This contrasts with what was found in the constant scenario, where peak motivation, peak fatigue, and work period duration all decreased as skill increased  (Fig.~\ref{fig:result_const_T}-C, D, E), and is distinct from the periodic changes in the stepwise scenario (Fig.~\ref{fig:result_3_panel}-A3, A4, A5). 

Both the stepwise and continuous scenarios can lead to sustained increase in skill learning and performance, but how far can one push the pace of task difficulty increase? We evaluated a special case in the continuous scenario where we increase the task difficulty at a very high rate (Fig.~\ref{fig:result_3_panel}-C). In this scenario, individuals give up after failing to keep up with the ever more challenging task (Fig.~\ref{fig:result_3_panel}-C). Notably, on short timescales, this special case is characterized by a substantial increase in both peak fatigue (Fig.~\ref{fig:result_3_panel}-C4) and the duration of resting periods (Fig.~\ref{fig:result_3_panel}-C5). These short-timescale dynamics suggest the task became too demanding, with the individual taking longer recover, without sufficient motivation to compensate for the escalating exertion.

\begin{figure}
    \centering
    \includegraphics[width = 0.8\textwidth]{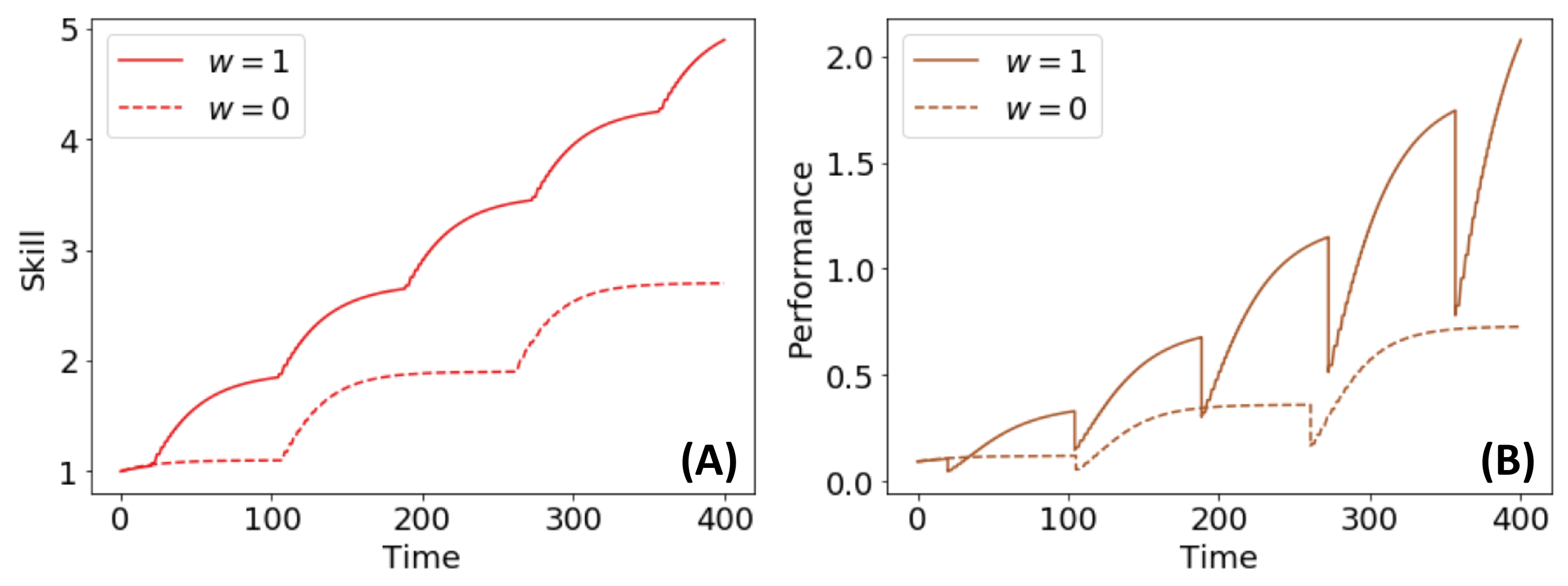}
    \caption{\textbf{Over the long run, learning-driven individuals achieve higher skill and performance than execution-driven individuals.} (A) Skill learning in individuals motivated by skill ($w = 1$; solid line) and execution ($w = 0$; dashed line). When individuals are learning-driven, they are quicker to attempt more difficult tasks, compared to execution-driven individuals. They thus achieve a higher level of skill. (B) Performance in individuals motivated by skill ($w = 1$; solid line) and execution ($w = 0$; dashed line). Since execution-driven individuals are slower to attempt more difficult tasks, they suffer fewer dips in performance than learning-driven individuals. The performance of learning-driven individuals, therefore, may repeatedly drop below the performance of execution-driven individuals. This short term deficit, however, is overcome by the learning-driven individuals' improved skill, so that eventually learning-driven individuals perform better than execution-driven individuals, even when the learning-driven individual is experiencing a dip in performance while the execution-driven individual is at their peak (e.g., Time = ~350).}
    \label{fig:result_w_comparison}
\end{figure}


\subsection{Individual differences: Learning-driven individuals overtake execution-driven individuals over the long run}
We next analyze the impact of individual differences in motivation on the long-term development of skill and performance. We manipulated whether individuals were primarily motivated by learning (i.e., rate of increase of skill) or execution. According to Eq.~\ref{eq:m}, learning-driven individuals (i.e., high $w$ value) derive proportionally more motivation from their skill development ($dS/dt$), while execution-driven individuals (i.e., low $w$ value) derive proportionally more motivation from their rate of execution ($E(t)$).

To capture the intuition that learning-driven individuals are quicker to abandon a too-easy task, while execution-driven individuals continue to optimize their current performance by persisting in a task they have nearly mastered, we set $\epsilon$---the threshold for increasing task difficulty in our stepwise scenario---to depend on whether individuals are motivated more by learning or execution: $\epsilon = \epsilon_{\text{min}} + k w$, where $\epsilon_{\text{min}} = 0.001$, and the slope $k = 0.05$. 

As shown in Fig.~\ref{fig:result_w_comparison}, early on in the stepwise scenario, learning-driven individuals perform worse than execution-driven individuals. This was due to the tendency for learning-driven individuals to increase task difficulty sooner than execution-driven individuals, and increases in task difficulty are accompanied by a dip in performance. This reflects the learning-driven individual's drive to seek out more challenging tasks, even at the cost of short-term dips in performance. However, the learning-driven individual's more frequent jumps eventually lead to higher skill than the execution-driven individual, and consequently, they outperform the execution-driven individual in the long run.




\section{Discussion}
Life-long learning involves non-linear change on multiple timescales. Motivation waxes and wanes from moment to moment, performance can rise and dip as new challenges are attempted, and skill can develop gradually over days or years. Our formal model offers a framework for relating these dynamics. In our model, the gradual development of skill emerges from short timescale fluctuations in motivation, fatigue, and effort. Performance can exhibit both gradual increases but also short-term dips, particularly when a new task or strategy is first attempted. The peaks and troughs of motivation and fatigue---and thus the amount of time individuals are able to sustain their effort---reflect individual differences in whether one is motivated by intrinsic growth (i.e., skill development) or extrinsic achievement (i.e., performance). By situating these dynamics within different training regimes---that is, different plans for increasing the task difficulty over time---we investigated how people can achieve life-long learning or fail to maintain their personal growth. 

\subsection{Challenge, motivation, and the elusive state of flow}
In our model, individuals are motivated in part by the gap between their current skill and the difficulty of the task ($T-S$). As a result, peak motivation (the maximum motivation reached during a work session) is highest when the task is appropriately difficult and decreases as individuals approach mastery. This early period of maximal motivation, moreover, is accompanied by longer periods of sustained effort---despite high levels of accumulated fatigue. In other words, when the task is sufficiently challenging, and skill is sufficiently high, then individuals experience high motivation, the capacity to work for extended periods of time, and the capacity to sustain their efforts in the face of high fatigue (Fig. \ref{fig:result_const_T}-C, D). These are all hallmarks of the state of flow, an elusive state of optimal engagement \cite{csikszentmihalyi1990flow}. 

While flow has traditionally been discussed as a phenomenon that emerges and disappears on short timescales---commonly emerging during a single session of work---our model connects this phenomenon to the long timescale dynamics of skill development, performance, and task selection. When task difficulty increases in discrete jumps, for instance, individuals experience brief periods of flow-like engagement, which disappear as their skill catches up to the demands of the task (Fig.~\ref{fig:result_3_panel}-A). When the difficulty of the task is increased continuously, however, flow-like engagement can be sustained and even increased over time (Fig.~\ref{fig:result_3_panel}-B).

\subsection{Leaps, plateaus, and sustained learning}
The emergence of flow-like engagement in our model depends on the source of an individual's motivation. When individuals are intrinsically motivated by learning (i.e., $w$ close to $1$), they sacrifice short-term gains in performance for the opportunity to improve their skill (Fig.~\ref{fig:result_w_comparison}). This leads to frequent dips in performance. Over time, however, the benefits of learning outweigh the costs of these performance dips, a phenomenon that has been described empirically in skill acquisition ranging from the discovery of the Fosbury Flop technique in the sport of high jump, to the development of prodigious memory for a span of random digits (\cite{gray2017plateaus}). Since performance dips are followed by periods of increased challenge and flow-like engagement, our model captures the notion of the so-called autotelic personality type, characterized by a disposition to actively seek challenges and flow experiences (\cite{baumann2021autotelic}). The long-timescale dynamics of learning thus depend on the moment-to-moment dynamics of motivation and task selection. 

Maximizing long-timescale learning, however, requires a careful balance. For individuals to experience sustained, long-term learning, they must engage in challenges above their current skill level, but only to a degree. If individuals choose to maximize their current performance rather than seek out new challenges (such as the execution-driven individuals in Fig.~\ref{fig:result_w_comparison}), then their long-term learning will progress more slowly. But if task difficulty is increased too quickly, then learning cannot keep up and eventually an individual will cease to learn, perform, and work (Fig.~\ref{fig:result_3_panel}-C). When the increase in learning is carefully calibrated to an individual's ability (Fig.~\ref{fig:result_3_panel}-A and B), then learning and peak motivation can both grow over long periods \cite{chow2013effective,son2006metacognitive,headrick2015dynamics}.

The decision to attack entirely new tasks, or the discovery of a more difficult but potentially more effective approach to an existing task, can produce learning curves that consist of piece-wise power laws (e.g., \cite{donner2015piecewise}). Here, our model reproduces this long-timescale dynamic when the task difficulty is increased in discrete jumps (Fig.~\ref{fig:result_3_panel}-A), with each period of decreasing returns followed by a period of renewed rapid growth. In our model, however, each period of decreasing returns actually follows an \emph{exponential} rather than power-law learning curve, in line with work showing that individual learning curves follow an exponential form \cite{estes1956problem,  heathcote2000power, anderson1997artifactual}. As we show analytically in the Appendix, this exponential decay in the rate of learning emerges from our assumption that the rate of learning depends only on relative task difficulty (i.e., the difference between task difficulty and skill, $T-S$). In order for learning to follow a power law, the rate of learning must depend not only on the amount of current challenge but also on the passage of time itself (such as \cite{heathcote2000power}, which derives the same result). The difference between exponential and power laws of practice, therefore, reduces to whether the classic phenomenon of diminishing returns depends only on changes in skill (i.e., diminishing as skill approaches saturation), or on both skill and time (e.g., also diminishing with each attempt). We speculate, therefore, that the functional form of an individual's learning curve may depend on the individual and the task. In cases where the rate of learning decays with time (e.g., with an individual who just tries less over time), then learning should follow a power law; in cases where the rate of learning depends only on the current challenge, then learning should follow an exponential curve. 

\subsection{Conclusion and contributions}
Skill mastery is the outcome of processes that operate on widely varying timescales. Motivation and fatigue can change rapidly in response to context and activity, decisions to work or rest are based on fluctuations in motivation and fatigue, while long-term improvements in skill and performance reflect days or years of work. 

Our work offers two contributions. First, our model offers a formal mathematical framework to study the interaction of these timescales, and to understand the multiscale dynamics of learning as an integrated dynamical system. Short-timescale dynamics shape long-timescale dynamics. The functional form of the long-timescale learning curve, for instance, depends on the instantaneous rate of learning. The influence goes in the other direction, too. Different long-timescale regimes for increasing task difficulty, for instance, produce different short-timescale dynamics of motivation and fatigue. While common sense suggests that there must be interactions between processes on very different timescales, our model spells out some of those potential interactions. 

Second, our work brings into conversation different research areas that are typically investigated in isolation. The literature on flow states remains focused on subjective reports, and is seldom connected to the mathematical psychology of learning curves. Research on life-long mastery is seldom connected to the small improvements possible during a session of practice. The mastery achieved by virtuosos such as the artist Picasso or the athlete Muhammad Ali is treated separately from the modest achievements of regular people. Nevertheless, learning occurs on all timescales, and is common to all humans. We offer here a unified framework that connects these different perspectives. We hope our work encourages scientists working in separate disciplines to start a collective conversation about the complex, multiscale dynamics of skill learning. 

    
\section*{Methods: Details of the mathematical model}
We implement our model as a dynamical system consisting of a system of ordinary differential equations. We divide the description of the dynamical system into two components: three state variables that change on the long timescale, and another three that change on short timescale. Fig.~\ref{fig:model}-A illustrates the interactions between these short- and long-timescale variables. See Table~\ref{tab:model_paras} for a summary of key model parameters on various time scales. 

\subsection*{Long timescale}

The three key long-timescale variables are $S(t)$, the skill level of an individual at time $t$, $E(t)$, the rate of execution, or how much work an individual is getting done at time $t$, and the task difficulty, $T(t)$. $S$, $E$, and $T$ all take continuous, non-negative values. The distinction between skill and execution captures the fact that people often learn best and perform best at different times \cite{soderstrom2015learning}. One performs best when executing tasks that are easy, relative to one's skill level, while one develops skills when working on tasks that are more difficult (but not too much more difficult) than one's current skill level \cite{akizuki2015measurement,guadagnoli2004challenge}. We calculate an individual's performance as the product of their current skill and current execution, $S(t)E(t)$, since performance is jointly affected by skill level and rate of execution. 

\subsubsection*{Skill development and execution}


In the model, skill improvement is a function of \textit{relative task difficulty}, which is the difference between task difficulty and one's skill level $T-S$. In other literature, the same variable is also referred to as functional task difficulty \cite{akizuki2015measurement}. Specifically, skill improves when the task is harder than one's current skill level ($x>0$), but only to an extent. If a task is too easy or too difficult, the individual working on it does not develop skill \cite{guadagnoli2004challenge, akizuki2015measurement}. Put mathematically, the dynamics of skill are given by: 
\begin{equation}\label{eq:s}
    \frac{dS}{dt} = p f_S(T-S)\;, 
\end{equation}

where, for simplicity, $f_S$ takes the shape of a triangle, peaking at value $x_m$ (where $x_m>0$), and is visualized in Fig.~\ref{fig:model}-D. The function can be expressed in the following piecewise form: 
\begin{equation}
f_S(x) = \begin{cases} 
     0 & \text{if } x < 0\;,\\
     y_m x/x_m & \text{if } 0 \leq x \leq x_m  \;, \\
      y_m (-x + 2 x_m)/x_m & \text{if }  x_m < x \leq 2 x_m \;,\\
      0 & \text{if }  x > 2 x_m \;, 
\end{cases}
\end{equation}\label{eq:fs_1}

where $y_m$ is the height of the triangle function. 

Note that this approach, unlike power-law or exponential learning curves,  allows us to model the period of skill acquisition when the initial task is much too difficult for the learner, given their current skill. During this period, learning is laborious and slow, with little skill acquisition occurring even with extended effort. Think of the young child who first encounters chess before they are sufficiently mature to even remember all the rules: repeated encounters with the game will produce little improvement in their ability to play chess. This period is often ignored in empirical work on skill acquisition, because participants and tasks are intentionally matched so the task is approachable. However, in many real-world scenarios, an individual encounters a task that is too difficult to even start---or if starting is possible, then it is initially quite laborious. In our modeling framework, this period corresponds to the right-hand side of the learning function where relative task difficulty is very high.

Second, note that our approach is agnostic to the exact functional form of the learning curve. We show in the Appendix that the function $f_S$, which governs the rate of learning as a function of the relative task difficulty, can be modified to generate learning curves that follow, exactly, either an exponential or a power law curve. Here, we choose the simplest functional form that gives diminishing returns on effort: a piecewise linear relationship between relative task difficulty and rate of learning. This allows us to capture the qualitative features of empirical learning curves with fixed tasks, while also allowing us to model more complex scenarios involving variable task difficulty (e.g., when task difficulty is simultaneously changing with time or adapting to skill).

Similarly, the dynamics of execution are given by, 
\begin{equation}
    E(t) = p f_E(T-S)\;,
\end{equation}
where the shape of the function $f_E$ is sketched in Fig.~\ref{fig:model}-D. Rate of execution $E(t)$ is a constant when relative task difficulty is negative ($T-S<0$), and decays and eventually settles to $0$ when relative task difficulty increases ($T-S> 0$). The function can be expressed in the following piecewise form: 
\begin{equation}
f_E(x) = \begin{cases} 
      y_m & \text{if } x < 0 \;,\\
      y_m (-x + x_m)/x_m & \text{if }  0 \leq x \leq x_m \;,\\
      0 & \text{if } x > x_m.
\end{cases}
\end{equation}\label{eq:fa_1}

More informally, this says that execution is at a maximum when the demands of the task are fully met by one's skill, but execution decreases linearly as the deficit between task difficulty and skill gets larger. 

An individual's performance on a task reflects both their skill and the rate of execution. Performance is highest when they are both highly skilled and executing the task at the maximum rate; it is lower for individuals with lower skill, individuals who are struggling to execute the task, or both. We thus measure an individual's performance as the product of their skill and their rate of execution, $S(t)E(t)$. 

\subsubsection*{Task Difficulty}
Unlike skill ($S$) and execution ($E$), which are solely governed by the internal dynamics of the dynamical system, our third long-timescale variable, task difficulty ($T$), can be treated as an external input variable. As a result, our modeling framework allows us to explore how different ways of increasing task difficulty affect the dynamics of long-timescale learning and short-timescale engagement. 

Specifically, we evaluated several different training regimes outlined in the main text: task difficulty remains unchanged (constant scenario), increases in a discrete fashion once skill catches up with task difficulty (stepwise scenario), and increases in a continuous fashion (continuous scenario).

\subsection*{Short timescale}
On the short timescale, three key variables governing the working behavior are motivation ($M$), fatigue ($F$), and work ($P$). 

Motivation refers to one's current willingness or desire to work on a task \cite{brehm1989intensity}. For simple tasks that are primarily physical, this may be driven by levels of the neurotransmitter dopamine \cite{martin2018mental}; for more complex or conceptual tasks, motivation intensity is likely a composite of different psychological and physiological cues. Here, we remain agnostic about the specific physiological underpinning of motivation. We model motivation, $M$, as a continuous variable taking non-negative values.

Fatigue is a sense of cumulative exertion or tiredness. For some simple tasks that are primarily physical, the sense of fatigue may be driven by simple physiological processes, such as the accumulation of cerebral adenosine \cite{martin2018mental}; for more complex or conceptual tasks, much like for motivation, the sense of fatigue is likely a composite of different psychological and physiological cues. Here we also remain agnostic about the specific physiological underpinning of fatigue. We model fatigue, $F$, as a continuous variable taking non-negative values.


\subsubsection*{Motivation}
We assume the level of motivation is affected by two processes. The first is that, in the absence of other influences, motivation will restore to a baseline ($M_0$) \cite{coa2016baseline,ondersma2009motivation}. The second is that motivation increases when an individual is learning or executing a task \cite{atkinson1957motivational,anderman2020achievement}. Mathematically, these dynamics are expressed as 

\begin{equation}\label{eq:m}
    \frac{dM(t)}{dt} = c_1 (M_0 - M(t))M(t) + c_2 \left[w \frac{dS(t)}{dt} + (1-w) E(t)\right]\;,
\end{equation}
where the first term reflects the process of returning to baseline, and the second term captures the increase in motivation due to learning and execution. The variables $c_1$ and $c_2$ are constants weighting the contribution of these two processes. As introduced above, $S(t)$ denotes skill and $E(t)$  denotes execution of an individual at time $t$. The relative influence of learning and execution on motivation is determined by parameter $w$, which takes values between $0$ (only execution matters) and $1$ (only learning matters). 

\subsubsection*{Fatigue}
The level of fatigue is also affected by two processes. The first is that when one is working, the level of fatigue increases \cite{massar2010manipulation}. Here, we use the simplest form of this process, a linear increase. The second is that when one is resting, the level of fatigue recovers to baseline, zero. Here, we draw on the literature on recovery (e.g., \cite{konz1998work}), which has found that recovery tends to involve an exponential decay in fatigue. These dynamics of accumulating and decaying fatigue are mathematically integrated into the following piecewise function, 

\begin{equation}\label{eq:f}
    \frac{dF(t)}{dt} =
\begin{cases} 
      c_3  &  \text{if} \;p(t)= 1 \;,\\ 
     -c_4 F(t) & \text{if} \;p(t) = 0 \;, 
\end{cases}
\end{equation}

where $p$ indicates whether the individual is working ($p= 1$) or resting ($p = 0$). This differential equation formulation is equivalent to a linear increase in time when working, and exponential decay when resting. $c_3$ and $c_4$ are constants that set the timescales of the changes. 

\subsubsection*{Work}
An individual's decision to work or rest depends on their motivation and fatigue. An individual will switch to working from resting when they have sufficient motivation, relative to their fatigue---that is, when $M-F$ increases across a critical threshold, $\theta$.  They stop working when their fatigue is equal to or greater than their motivation---that is, when $M-F$ is smaller than or equal to zero. A sketch of this dynamic is shown in Fig.~\ref{fig:model}(C). In discrete time, the function $p$ can be expressed as 

\begin{equation}\label{eq:p}
p(x, t + \Delta t) = \begin{cases} 
      0 & \text{if } p(x, t) = 0 \text{ and } x < \theta \text{ , or } p(x, t) = 1 \text{ and } x \leq 0 \;,\\
      1 & \text{if } p(x, t) = 0 \text{ and } x \geq \theta \text{ , or } p(x, t) = 1 \text{ and } x > 0 \;.
\end{cases}
\end{equation}

\begin{table}
\centering
\caption{Summary of key model variables and parameters}
\begin{tabular}{ | c | p{8cm} | c | }
 \hline
 \hline
 \multicolumn{3}{|c|}{Long Timescale Variables} \\
 \hline
 Variable & Meaning & Range\\
 \hline
 $S$ & Individual's skill level & $[0, \infty)$ \\
 $E$ & Individual's rate of execution on task & $[0, \infty)$  \\
 $T$ & Task difficulty & $[0, \infty)$ \\
 \hline
 \hline
 \multicolumn{3}{|c|}{Short Timescale Variables} \\
 \hline
 Variable & Meaning & Range\\
 \hline
 $M$ & Individual's current motivation & $[0, \infty)$ \\
 $F$ & Individual's current fatigue  & $[0, \infty)$ \\
 $P$ & Whether individual is working (1) or resting (0) & $\{0,1\}$ \\
 \hline
 \hline
 \multicolumn{3}{|c|}{Parameters} \\
 \hline
 Parameter & Meaning & Range\\
 \hline
 $w$ & Whether individual is more driven by the accumulation of achievements (0) or skills (1)  & $[0, 1]$ \\
 \hline
 \hline
\end{tabular}
\label{tab:model_paras}
\end{table}

\subsection*{Computational simulations} \label{sec:methods_comp}
In order to generate predictions from the model, we numerically compute the solution for the ordinary differential equation system Eqs.~\ref{eq:s}, \ref{eq:m}, \ref{eq:f}, and \ref{eq:p} in the Python 3 programming language. 

We numerically evaluated three different training regimes outlined in the main text:  Task difficulty (1) remains unchanged (constant scenario), (2) increases in discrete jumps once skill catches up with task difficulty (stepwise scenario), and (3) increases in a continuous fashion (continuous scenario).

For the constant scenario (Fig.~\ref{fig:result_const_T}, we set task difficulty to a constant $T = 1.1$. The model parameters used were $M_0 = 1$, $w = 0.8$, $x_m = 1.5$, $y_m = 0.2$, $c_1 = 1$, $c_2 = 25$, $c_3 = 0.5$, and $c_4 = 0.5$. Initial conditions were $S(0) = 1$, $E(0) = 0 $, $M(0) = 1.5$, $F(0)= 1.4$, $P(0) = 0$, $dS/dt(0) = 0$. 

For the stepwise scenario (Fig.~\ref{fig:result_3_panel}-A) we used the same parameters and initial conditions as those for the constant scenario. The only difference between the two simulations is the change in task difficulty. We set task difficulty to increase by a fixed increment ($0.8$) whenever $T-S$ became smaller than a threshold, $\epsilon$, which is set to $0.04$ in this simulation. 

For the continuous scenario (Fig.~\ref{fig:result_3_panel}-B and C), we used parameters $M_0 = 1$, $w = 0.8$, $x_m = 1.5$, $y_m = 0.1$, $c_1 = 1$, $c_2 = 100$, $c_3 = 1$, and $c_4 = 1$. The simulations used the same initial conditions as the two simulations above. For the case of sustained learning (Fig.~\ref{fig:result_3_panel}-B), task difficulty was $T(t) = 0.01t$. For the case that leads to unsustainable learning (Fig.~\ref{fig:result_3_panel}-C), task difficulty was $T(t) = 0.03t$. 

For the simulation demonstrating differences between learning vs.~execution-driven individuals (Fig.~\ref{fig:result_w_comparison}), the parameters used were $M_0 = 1$, $x_m = 1.5$, $y_m = 0.1$, $c_1 = 1$, $c_2 = 25$, $c_3 = 1$, and $c_4 = 1$. For the learning-driven individual, $w = 1$, for the execution-driven individual, $w = 0$. The initial conditions were the same as the previous simulations. To capture the intuition that learning-driven individuals are quicker to abandon a too-easy task, while execution-driven individuals continue to optimize their current performance by persisting in a task they have nearly mastered, we set epsilon to depend on whether individuals are motivated more by learning or execution: $\epsilon = \epsilon_{\text{min}} + k w$, where $\epsilon_{\text{min}} = 0.001$, and the slope $k = 0.05$. 

The code used in the simulation is available online at \url{https://github.com/vc-yang/multiscale_learning_model}.



\bibliography{references}

\newpage
\section*{Appendix: Derivation of exponential and power-law learning curves} 
Psychologists have proposed two formal descriptions of the diminishing returns typically observed in learning curves: exponential and power law. Here we demonstrate that both these classic functional forms can follow naturally from our framework. For analytic tractability, we focus on the paradigmatic scenario of skill learning for a single fixed task, and ignore the short-timescale dynamics of motivation, fatigue, work, and rest. 

Recall that, within the model presented in the main text, skill increases when an individual is working ($p=1$) and as a function of relative task difficulty (i.e., difference between task difficulty $T$ and current skill $S$): 
\begin{equation}\label{eq:appendix_ds}
    \frac{dS}{dt} = p f_S(T-S)\;, 
\end{equation}
where the function $f_S$ captures qualitatively distinct phases of learning (sketched in Fig.~\ref{fig:model}-D):

\begin{equation}\label{eq:appendix_fs}
f_S(x) = \begin{cases} 
     0 & \text{if } x < 0\;,\\
     y_m x/x_m & \text{if } 0 \leq x \leq x_m  \;, \\
      y_m (-x + 2 x_m)/x_m & \text{if }  x_m < x \leq 2 x_m \;,\\
      0 & \text{if }  x > 2 x_m \;. 
\end{cases}
\end{equation}
Here, $2x_m$ is the width and $y_m$ is the height of this triangle function. 

In the first phase, one's skill is greater than the task difficulty ($T-S$ < 0) and no learning occurs ($dS/dt = 0$). In the second phase, the task is sufficiently challenging to induce learning, and skill learning increases linearly with relative task difficulty; in this phase, the learning rate will decrease dynamically as skill inches closer to task difficulty (i.e., diminishing returns). In the the third phase, task difficulty is greater than the optimal challenge ($T-S > x_m$), so the rate of learning declines with increasing difficulty. Finally, learning stops entirely if relative task difficulty is beyond a threshold ($T - S = 2x_m$).  

Here, we focus on the phase during which learning exhibits diminishing returns ($0 \leq T-S \leq x_m$), and ignore the dynamics of work and rest (i.e., set $p=1$). Assuming for simplicity that the maximum rate of increase of skill is 1 (i.e., $y_m=1$), and combining Eqs.~\ref{eq:appendix_ds} and \ref{eq:appendix_fs}, the the rate of change of skill is thus given by
\begin{equation}\label{eq:appendix_ds_simplify}
\frac{dS}{dt} = \frac{T-S(t)}{x_m}\;.
\end{equation}

First, we show analytically that this minimal model of learning gives rise to an exponential learning curve for skill $S$ as a function of time, $t$. 

Solving Eq.~\ref{eq:appendix_ds_simplify} gives 
\begin{equation}\label{eq:appendix_S}
    S(t)= T + \exp(- t/x_m) C\;,
\end{equation}
where $C$ is a constant determined by the initial condition of the system. If $S(0) = 0$, then $C= -T$, and the solution is $S(t) = T - T\exp(- t/x_m)$. Thus the dynamics of learning will exhibit \emph{exponential} diminishing returns toward the constant task difficulty $T$. 

Next, we show that our minimal model gives rise, analytically, to the classic ``power law of learning''  if the rate of learning depends not only on relative task difficulty (i.e., $T-S$) but also on the passage of time itself. We take Eq.~\ref{eq:appendix_S} and add a term that decays with time $t$:
\begin{equation}
\frac{dS(t)}{dt} =  \left(\frac{T - S(t)}{x_m}\right) \left(\frac{1}{t+1}\right)
\end{equation}

Integrating this equation, we get the learning curve: 
\begin{equation}
    S(t) = T + C \; (t+1)^{-1/x_m}\;,
\end{equation}
where $C$ is a constant determined by the initial condition of the system. If $S(0) = 0$, then $C= -T$, and the solution is $S(t) = T - T (t+1)^{-1/x_m}$. Thus the dynamics of learning will exhibit \emph{power-law} diminishing returns towards the constant task difficulty $T$. 

The difference between exponential and power laws of practice, therefore, reduces to whether the classic phenomenon of diminishing returns depends only on changes in skill (i.e., as skill approaches saturation), or on both skill and time.


\end{document}